# Unexplored photoluminescence from bulk and mechanically exfoliated few layers of $Bi_2Te_3$[†]

Bipin Kumar Gupta*[,1], Rabia Sultana[1, 2], Satbir Singh[1, 2], Vijeta Singh[1,2], Geet Awana[1,3], Anurag Gupta[1], Bahadur Singh[4], A. K. Srivastava[1], O.N. Srivastava[5], S. Auluck[1] and V.P.S. Awana*[,1]

[1]CSIR- National Physical Laboratory, Dr K S Krishnan Road, New Delhi, 110012, India

[2]Academy of Scientific and Innovative Research (AcSIR), CSIR-National Physical Laboratory Campus,

Dr K S Krishnan Road, New Delhi 110012, India

[3]Department of Physics and Astrophysics Delhi University, New Delhi-110007, India

[4]Department of Physics, Indian Institute of Technology, Kanpur, U.P.-208016, India

[5]Department of Physics, Banaras Hindu University, Varanasi, U.P. - 221005, India

**We report the exotic photoluminescence (PL) behaviour of 3D topological insulator $Bi_2Te_3$ single crystals grown by customized self-flux method and mechanically exfoliated few layers (18 ±2 nm)/thin flakes obtained by standard scotch tape method from as grown $Bi_2Te_3$ crystals. The experimental PL studies on bulk single crystal and mechanically exfoliated few layers of $Bi_2Te_3$ evidenced a broad red emission in the visible region from 600 nm – 690 nm upon 375 nm excitation wavelength corresponding to optical band gap of 2 eV. These findings are in good agreement with our theoretical results obtained using the *ab initio* density functional theory framework. Interestingly, the observed optical band gap is several times larger than the known electronic band gap of ~0.15 eV. The experimentally observed 2 eV optical band gap in the visible region for bulk as well as for mechanically exfoliated few layers $Bi_2Te_3$ single crystals clearly rules out the quantum confinement effects in the investigated samples which are well known in the 2D systems like $MoS_2$, $WS_2$, $WSe_2$, $MoSe_2$ etc for 1-3 layers.**

*****Correspondence tobipinbhu@yahoo.com (B.K.G.) and awana@nplindia.org(V.P.S.A.)**

Recently, topological insulators (TIs) have gained a huge attention among condensed matter physicists credited to their striking multifunctional properties.[1, 2] The most famous examples of 3D TIs are $Bi_2Te_3$, $Bi_2Se_3$ and $Sb_2Te_3$. These materials act as prototypical TIs because of their large intrinsic electronic band gap with a single Dirac-cone-like topologically protected metallic surface state inside the bulk energy gap. The interesting duality of this quantum materials viz., insulating from interior and conducting at the surface gives rise to many interesting properties with novel physics and potential for applications. Additionally due to the spin-momentum locked Dirac-cone surface states, TIs exhibit large un-saturating magneto resistance[3-5] and are appealing for applications in quantum information processing, magneto electric devices and next generation electronics/spintronics devices.[1,6-8] Most of the earlier studies on TIs focus on their mysterious quantum transport properties.[1-8] More recently however there has been a growing interest on their optical properties.[9-11] In particular, related to their nano-structuring and the confinement effects, it has been proposed that three dimensional (3D) TIs such as $Bi_2Te_3$, $Bi_2Se_3$ as well as two dimensional (2D) layered materials ($MoS_2$, $MoSe_2$ etc.) could be the potential material for optical applications.[12-15] Motivated by the interesting quantum properties of 3D TIs, in the present investigation, we focus on the photoluminescence (PL) studies of both bulk and mechanically exfoliated few layers of 3D TI $Bi_2Te_3$ single crystal. We observe a broad red emission in the visible region from 600 nm to 690 nm upon 375 nm excitation wavelength which corresponds to an optical band gap of 2eV. In order to ensure the origin and confirmation of PL results, we have performed the PL measurements for both bulk and few layers (18 ±2 nm) thin flakes of $Bi_2Te_3$ several times. This is in good agreement with our theoretical calculations. We have also explored in details about the topological surface states and possible inter band transitions of observed PL, using first-principles calculations. Our results show that $Bi_2Te_3$ as well other related TIs can act as ideal model systems for next generation optoelectronic device applications.

## Results

Figure 1 shows the single crystal XRD pattern of bulk $Bi_2Te_3$ indicating that the crystal is highly textured i.e. c- axis oriented showing all the diffraction peaks corresponding to the (00l) reflections of the rhombohedral crystal structure with space group R3m (D5). The left inset of Figure 1 represents the powder XRD pattern of bulk $Bi_2Te_3$ single crystal [Figure 1(a)]. The lattice parameters found are a = b = 4.3866(2) Å and c = 30.4978(13)Å. The right inset of Figure1 displays the temperature dependence of electrical resistivity at different applied magnetic fields for the synthesized $Bi_2Te_3$ single crystal having a temperature range of 5K to 50K [Figure 1(b)]. The applied magnetic field varies from 0 kOe to 100 kOe. Here, the value of resistivity increases with increase in temperature from 5K to 50K representing metallic behavior with and without applied magnetic field. Thus, the nature of the plot is in agreement to that observed in our earlier reported literature.[16] The extensive ongoing research activity worldwide on TIs inspired us to further explore the PL properties of bulk and few layers of $Bi_2Te_3$. As luminescent, insulators and semiconducting materials are widely studied, where PL originates from energy transfer from host to activator in the former and recombination of electron hole pair in the latter case. Energy transfer from insulators to conductors is also investigated in some composite systems.[17,18] Apart from composites, there isn't any single material reported so far, where energy transfer from insulating to metallic states region takes place. It is well established that TIs are electronic materials having a bulk band gap like an ordinary insulator but have protected conducting states on their edges or surface. These states are protected by spin orbit interaction (SOI) and time reversal symmetry.[1] These striking features of TIs make it more interesting to explore its optical behavior particularly photoluminescence nature. Previous studies on 3D TIs showed existence of dual band gaps; 0.3 eV corresponding to electronic band gap and 2.1 to 2.3 eV corresponding to optical band gap for the case of CVD grown $Bi_2Se_3$ nanoplatelets.[14] The existence of optical band gap in these materials give rise to a strong PL which we have investigated on the basis of experimental and theoretical calculations. It is well-known that $Bi_2Te_3$ has a layered crystal structure which consists of five atomic layers blocks, known as quintuple layer (QL).[16] Each QL contains five atomic layers ordered as Te-Bi-Te-Bi-Te. The bonding inside a QL is strong ionic-covalent type whereas QLs are separated by weak van der Waals forces.[1,2,8] The van der Waals gap in between the QLs makes $Bi_2Te_3$ cleavable as well as optically active.

The PL measurements are done for both bulk $Bi_2Te_3$ single crystal as well as for mechanically exfoliated few layers of $Bi_2Te_3$. The optical images of both cases are shown in Figures 2(a) and 2(b), respectively. The representative TEM images for bulk single crystal and few layers are shown in Figures 2 (c) and 2(d), respectively, which clearly differentiate between them. The SEM micrograph of $Bi_2Te_3$ as shown in Figure S1 (see Supplementary Information)clearly visualizes the layered structure. The TEM micrograph of bulk $Bi_2Te_3$ is shown in Figure S2 (see Supplementary Information) to further distinguish between bulk and exfoliated few layers $Bi_2Te_3$. The decrease in transparency is associated with increase in number of QLs. Further, well resolved lattice fringes obtained from HRTEM micrograph in Figure S3 (see Supplementary Information) shows the formation of $Bi_2Te_3$. The inter-planer spacing from the HRTEM micrograph comes out to be 0.21 nm. The SAED pattern with well indexed diffraction rings further confirms the lattice formation of $Bi_2Te_3$ as shown in Figure S4 (see Supplementary Information). Moreover, the atomic force microscopy (AFM) result confirms the thickness of mechanically exfoliated layers of $Bi_2Te_3$ is (18 ±2) nm as shown in Figure S5.

Angle resolved photoemission spectroscopy (ARPES) result shown in figure S6 depicts energy distribution map (EDM) measured at the zone center from $Bi_2Te_3$ at a sample temperature of 20 K using 70 eV photon energy and p-polarized light. Dirac point is located at a binding energy of 300 meV below the Fermi level as shown by the green arrow on the EDM. The ARPES data is taken at Elettra synchrotron light source. This result clearly shows that the as synthesized $Bi_2Te_3$ single crystal is topological in nature because the band dispersion in the vicinity of the Fermi level as shown in Fig. S6 is purely of the topological surface state (upper part of the Dirac cone) although the Dirac point is at around 300 meV below the Fermi level. However, there are several reports discussing the presence of bulk bands near the Fermi level that are generally induced by the crystal disorder.[19] Thus, the absence of bulk bands near the Fermi level in our system suggests a high quality of the crystal with minimum defects. Figure 3(a) shows the PL emission spectra of both bulk $Bi_2Te_3$ single crystal and mechanically exfoliated few layers of $Bi_2Te_3$ upon excitation at 375nm marked in black and red color respectively. The emission spectrum of bulk $Bi_2Te_3$ single crystal shows a strong red emission at 621.32 nm (~2 eV). Also, several other closely spaced spectral peaks appears in the red region from ~ 630 nm to 700 nm. Similar peaks were also observed in the case of mechanically exfoliated few layers of $Bi_2Te_3$ in the same region. However, significant enhancement in PL intensity is observed in the case of few layers of $Bi_2Te_3$ because of the decreasing number of QL which induces the efficient photon emission from few layers of $Bi_2Te_3$ due to internal transitions taking place in the sub-levels of this system. Accordingly, the optical penetration loss effect depends on the sample thickness. The present experimentally observed PL emission centered about 2 eV differs from that observed in the range of 2.1 eV -2.3 eV for $Bi_2Se_3$ in another study.[14] The obvious reason could the difference in two TI systems with different morphologies and dimensions having different inter band transitions which enable different optical band gaps.

The few layers of $Bi_2Te_3$ are mechanically exfoliated from the bulk single crystal using the standard scotch – tape method[20-22] and hence, their thickness is considerably reduced to few QLs in comparison to the bulk $Bi_2Te_3$ single crystal where efficient photon emission is possible without the major peak shift like other 2D materials. In our investigated systems ($Bi_2Te_3$), we expostulate and understand the enhancement in PL emission is due to the higher photon emission into the thin flake of $Bi_2Te_3$ via Förster mechanism. More specifically, reduction of volume to surface area provides better photon energy emission due to internal transitions occurring in the sub-levels of the few layers of mechanically exfoliated $Bi_2Te_3$ without much of non-radiative loss like in the bulk system. Generally, PL may arise from such system due to defect or impurity present in crystal structure. The exact mechanism of the photon emission is however complex and requires further analysis. To further ensure the PL experiment and origin of PL, the absorbance analysis were performed to correlate with theoretical estimated data of mechanically exfoliated few layers of $Bi_2Te_3$. The reflectance spectrum of mechanically exfoliated few layers of $Bi_2Te_3$ is shown in Figure S7 (see Supplementary Information). Figure 3(b) depicts the absorbance spectrum of mechanically exfoliated few layers of $Bi_2Te_3$ calculated using the formula $A = Log_{10} 1/R$, where A and R represents the absorbance and reflectance respectively. Here, we noticed strong absorption peaks at 576 nm and 663 nm which have good agreement with the observed PL peaks at 620 and 650 nm. Moreover, few more peaks are observed similar to the theoretical calculation of partial density of states (PDOS) [Figure 4(a-b)]. In order to develop a better understanding of the underlying mechanism for observed PL, we employ density functional theory based calculations, using both the WIEN2k[23] as well as VASP[24] suite of codes.

The generalized gradient approximation (GGA) is used to include exchange correlation effects.[25] Our calculations are performed with the inclusion of spin-orbit coupling (SOC). We have used 6 QLs to explicitly model few layers of $Bi_2Te_3$. The band structure of a few layers (6 QLs) of $Bi_2Te_3$ with the inclusion of SOC is displayed in Figure 3(c) where the shaded part arises from the bulk. Our results show an insulating energy spectrum for bulk $Bi_2Te_3$ whereas its few layers have additional Dirac-cone-like conducting surface states inside the bulk energy gap. These results clearly demonstrate that $Bi_2Te_3$ is a topological insulator with different electronic states inside the bulk and on the surface and therefore, it is possible to have more photon emissions from such thin layers of $Bi_2Te_3$ due to the internal transitions of the sub-levels in the present TI system.

To further explore the energy levels electronic transitions responsible for PL emission, we show in Figure 4 (a-b) the partial density of states (PDOS) for bulk with the inclusion of SOC. It should be noted that PDOS of few layers of $Bi_2Te_3$ has significant change in comparison to the bulk, which is consistent with the experiments, except the appearance of small DOS at Fermi-level which is due to the presence of topological surface states. The PDOS profile clearly shows the existence of dominant electronic transitions between Te-p to Bi-p states. These electronic transitions have energies of ~1.7 eV and ~2 eV and are in substantial agreement with the observed PL peaks noted in Figure 3(a). We have also found additional electronic transitions at higher energies and are not shown here as they lie away from the range of our measured PL spectra. Alternatively, we have calculated the joint density of states (JDOS) which does not have any broadening. The JDOS shown in Figure 4(b) clearly identifies two dominant peaks at ~1.7 eV and ~2 eV, in qualitative agreement with our experimental observations except for the splitting size. This can be understood because of well known limitation of DFT in underestimating the gap. Hence a quantitative agreement with observed experimental is not possible.

**Discussion**

In conclusion, we report the optical band gap of bulk single crystals and mechanically exfoliated few layers of $Bi_2Te_3$ of around 2 eV, which is greater than its electronic band gap (~ 0.15eV).This is consistent with our first-principles calculations. The experimentally observed 2 eV optical band gap in the visible region for bulk as well as for mechanically exfoliated few layers of $Bi_2Te_3$ clearly rules out the quantum confinement effects in the investigated samples which are the new paradigm shifts from other 2D systems like $MoS_2$, $WS_2$, $WSe_2$, $MoSe_2$ etc for 1-3 layers. Our results provide a stepping stone for further exploration of topological insulators like $Bi_2Te_3$ for next generation optoelectronic, photonic, and spintronics applications.

**Methods**

The synthesis of bulk $Bi_2Te_3$ single crystal used for this study were carried out by self-flux method via the solid state reaction route.[16] High purity (99.999%) Bismuth (Bi) and Telluride (Te) powders were accurately weighed according to the stoichiometric ratio of 2:3 and mixed thoroughly using mortar and pestle inside a glove box filled with argon gas. The homogeneous mixture was then pressed into a rectangular pellet with the help of hydraulic press (40Kg/cm$^2$) and vacuum sealed (10$^{-3}$Torr) in a quartz tube. Further, the quartz tube was kept inside a tube furnace and heated to 950°C (2°C/min) for 12 hours. The furnace was then allowed to cool very slowly to 650°C (1°C/30min) after which it was switched off. The resultant crystal obtained was shiny and silver in color.

**Measurements and Characterizations.** In order to investigate the phase structure of the synthesized sample X-ray diffraction (XRD) was performed using Rigaku X-ray diffractometer with Cu-K$\alpha_1$ radiation ($\lambda$= 1.5418 Å). The temperature dependent resistivity measurements were carried out using a quantum design 14Tesla Physical Property Measurement System (PPMS). The thickness of the few layers of mechanically exfoliated $Bi_2Te_3$ was investigated by atomic force microscope (AFM, Model no. NT-MDT Solver Scanning probe Microscope). The absorbance measurements were carried out using Avantes spectrometer with AvaLight-DH-S-BAL balanced power source. For PL measurements, mechanically exfoliated[20-21] few layers of $Bi_2Te_3$ were obtained using the standard scotch tape method.[22] However, the PL measurements were done using WITec alpha 300R+ confocal PL microscope system using 375 nm diode laser as a source of excitation. In order to prevent the damage of bulk $Bi_2Te_3$ single crystal and few layers of $Bi_2Te_3$, caution was taken and the power setting for the excitation laser was maintained to lower values. Further, the surface morphology Transmission Electron Microscopy (TEM) was performed using Technai G20-twin operating at 200kV. The ARPES data istaken at Elettra synchrotron light sourceat a sample temperature of 20 K using 70 eV photon energy and p-polarized light. In order to understand the measured PL spectrum we present density functional theory based calculations of the optical properties of $Bi_2Te_3$ for bulk and few layer structures.

**FIGURE CAPTIONS:**

**Figure 1. (color online)** X-ray diffraction pattern for $Bi_2Te_3$ single crystal, Inset (a) shows the Rietveld fitted room temperature XRD pattern for powder $Bi_2Te_3$ crystal. (b) Temperature dependence of electrical resistivity at different magnetic fields for $Bi_2Te_3$ single crystal.

**Figure 2**. **(color online)** (a) Optical image of bulk $Bi_2Te_3$ single crystal. (b) Optical image of scotch tape taken few layers of $Bi_2Te_3$. (c) TEM image of bulk $Bi_2Te_3$ single crystal. (d) TEM image of few layers of $Bi_2Te_3$.

**Figure 3**. **(color online)** (a) PL spectrum of $Bi_2Te_3$ single crystal and mechanically exfoliated few layers. Green and red colors identify spectrum of bulk and few layers, respectively. (b) Absorbance spectrum of mechanically exfoliated few layers of $Bi_2Te_3$. (c) A few layers band structure of $Bi_2Te_3$ with the inclusion of spin-orbit coupling along high symmetry directions in the (001) surface Brillion zone. Note that we have taken 6 QLs in our calculations as a representative case for few layers $Bi_2Te_3$. The shaded part refers to the bulk states.

**Figure 4.** (color online) (a) Partial density of states (PDOS) and (b) Joint density of states (JDOS) for bulk $Bi_2Te_3$ with the inclusion of spin-orbit coupling. Te1 and Te2 in (a) represent Te atoms at the edge and center of QL, respectively.

**Acknowledgements**

Authors from CSIR-NPL would like to thank their Director NPL India for his keen interest in the present work. This work is financially supported by DAE-SRC outstanding investigator award scheme on search for new superconductors. Rabia Sultana thanks CSIR, India for research fellowship and AcSIR-NPL for Ph.D. registration. SA would like to thank the High Performance Computing (HPC) facilities at Intra-University Accelerator Centre (IUAC) at New Delhi, Indian Institute of Mathematical Sciences (IMSC) at Chennai, CSIR- 4PI at Bengaluru, University of Hyderabad (UoH) in Hyderabad and Physics Department Indian Institute of Technology at Kanpur.

**Author Contributions**

B. K. G. and V.P.S. A. conceived the concepts of the research. R.S., G.A. and A. G. designed and synthesized the samples as well as performed the resistivity vs temperature measurements. S.S. performed the AFM and PL measurements of sample. A.K.S. performed the transmission electron microscopy of the sample. V.S., B.S. and S.A. performed the theoretical analysis. B. K. G., V.P.S.A., R.S., O.N.S. and S.A. wrote the manuscript and analysed the data.

# Additional information

**Supplementary information** accompanies this paper at http://www.nature.com/scientific reports.

Competing financial interests: The authors declare no competing financial or non-financial conflicts of interest.

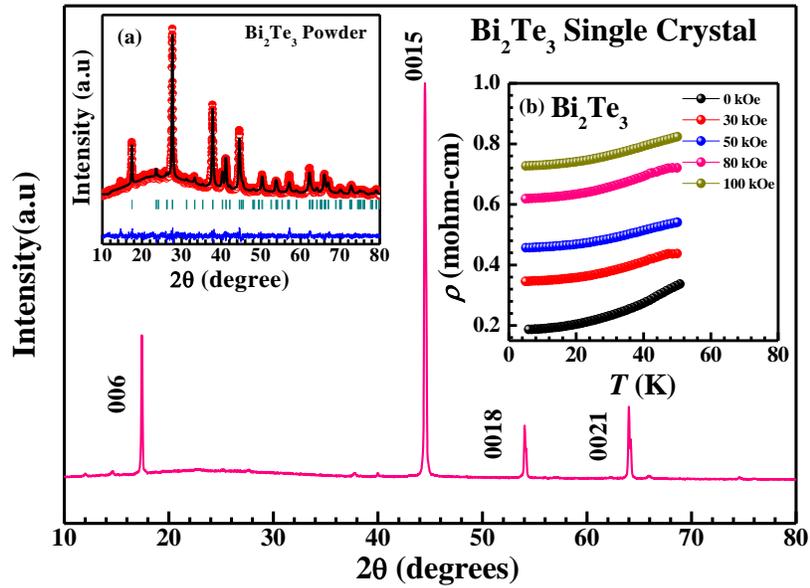

**Figure 1. (color online)** X- ray diffraction pattern for Bi$_2$Te$_3$ single crystal. Inset (a) shows the rietveld fitted room temperature XRD pattern for powder Bi$_2$Te$_3$ crystal. (b) Temperature dependence of electrical resistivity at different magnetic fields for Bi$_2$Te$_3$ single crystal.

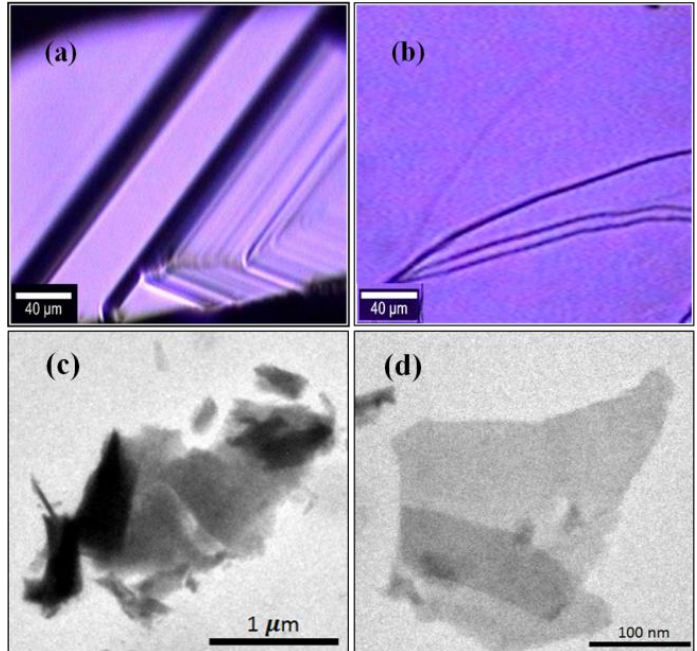

**Figure 2.(color online)** (a) Optical image of bulk Bi$_2$Te$_3$ single crystal. (b) Optical image of scotch tape taken few layers of Bi$_2$Te$_3$. (c) TEM image of bulk Bi$_2$Te$_3$ single crystal. (d) TEM image of few layers of Bi$_2$Te$_3$.

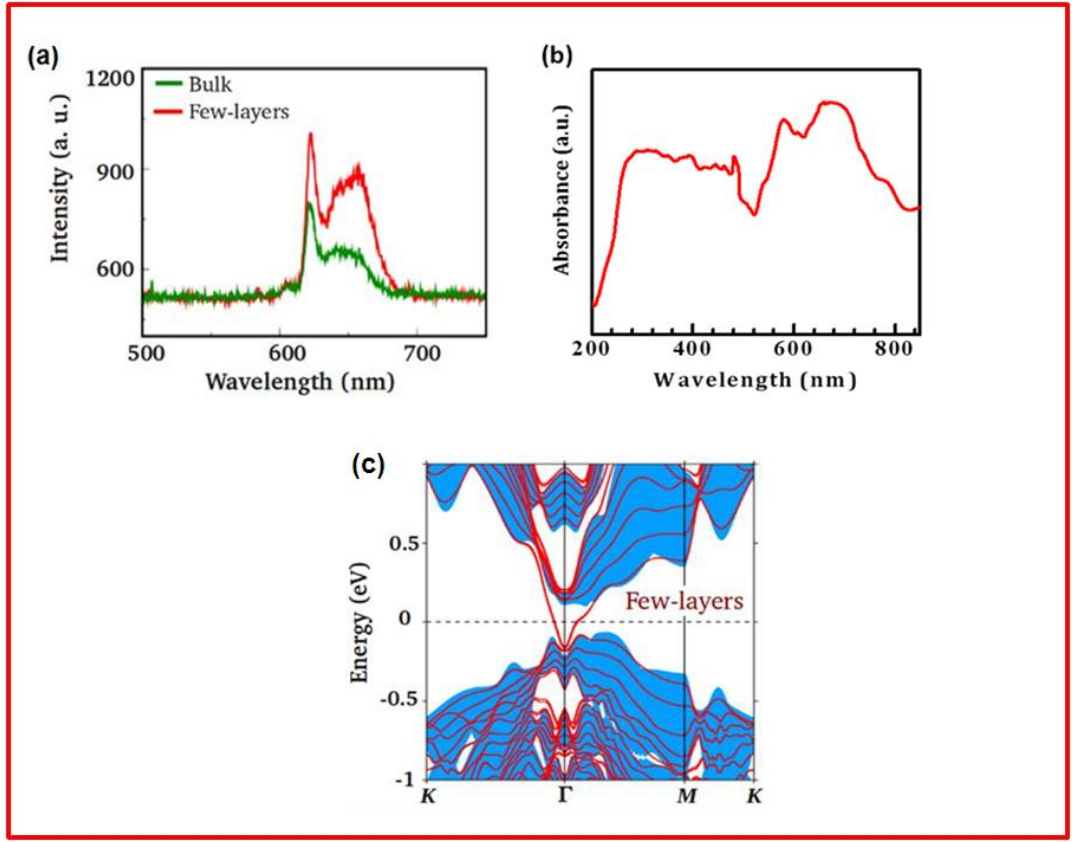

**Figure3**. **(color online)** (a) PL spectrum of $Bi_2Te_3$ single crystal and mechanically exfoliated few layers. Green and red colors identify spectrum of bulk and few layers, respectively. (b) Absorbance spectrum of mechanically exfoliated few layers of $Bi_2Te_3$. (c) A few layers band structure of $Bi_2Te_3$ with the inclusion of spin-orbit coupling along high symmetry directions in the (001) surface Brillion zone. Note that we have taken 6 QLs in our calculations as a representative case for few layers $Bi_2Te_3$. The shaded part refers to the bulk states.

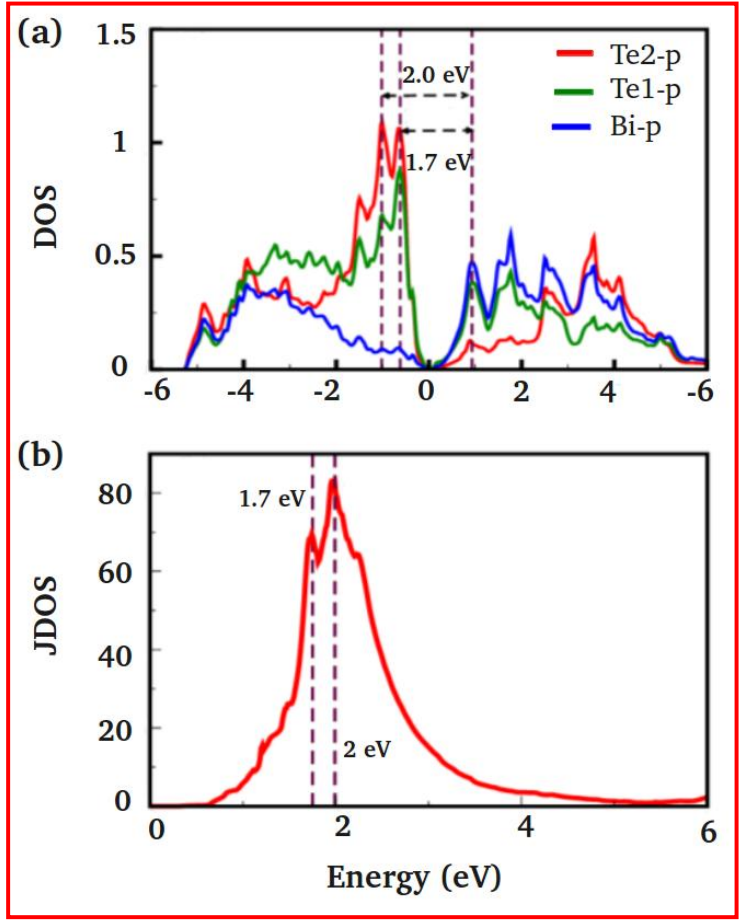

**Figure 4. (color online)** (a) Partial density of states (PDOS) where, Te1 and Te2 represent Te atoms at the edge and the center of QL, respectively and (b) Joint density of states (JDOS) for bulk $Bi_2Te_3$ with the inclusion of spin-orbit coupling.

# Supplementary Information

## Unexplored photoluminescence from bulk and mechanically exfoliated few layers of $Bi_2Te_3$

Bipin Kumar Gupta[*,1], Rabia Sultana[1, 2], Satbir Singh[1, 2], Vijeta Singh[1,2], Geet Awana[1,3], Anurag Gupta[1], Bahadur Singh[4], A. K. Srivastava[1], O.N. Srivastava[5], S. Auluck[1] and V.P.S. Awana[*,1]

[1]*CSIR- National Physical Laboratory, Dr K S Krishnan Road, New Delhi, 110012, India*

[2]*Academy of Scientific and Innovative Research (AcSIR), CSIR-National Physical Laboratory Campus, Dr K S Krishnan Road, New Delhi 110012, India*

[3]*Department of Physics and Astrophysics Delhi University, New Delhi-110007, India*

[4]*Department of Physics, Indian Institute of Technology, Kanpur, U.P.-208016, India*

[5]*Department of Physics, Banaras Hindu University, Varanasi, U.P. - 221005, India*

**\*Correspondence to  bipinbhu@yahoo.com (B.K.G.) and awana@nplindia.org(V.P.S.A.)**

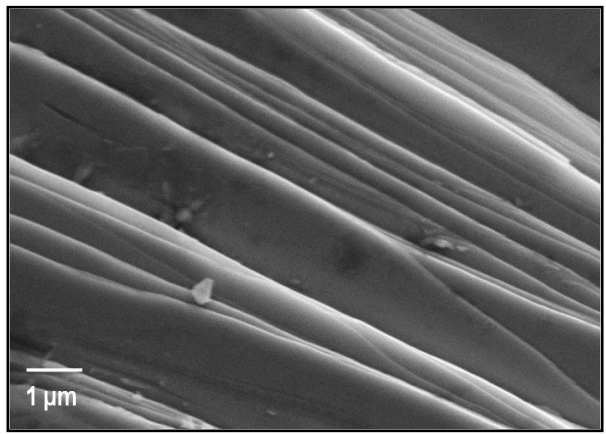

**Figure S1**. SEM micrograph of $Bi_2Te_3$ clearly shows the layered structure.

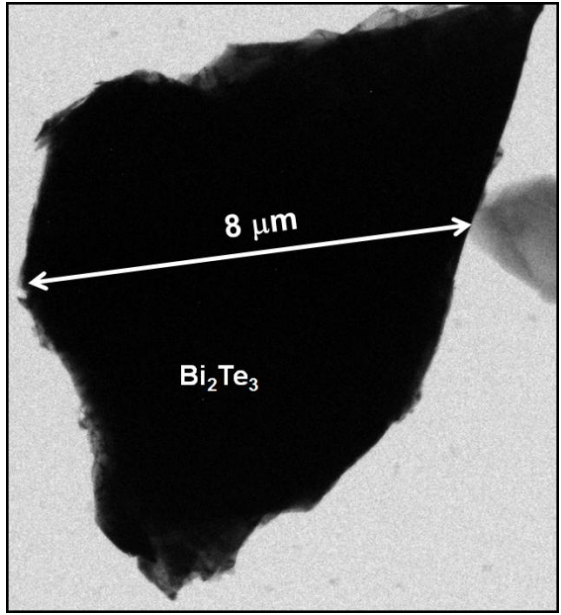

**Figure S2.** TEM micrograph of bulk $Bi_2Te_3$

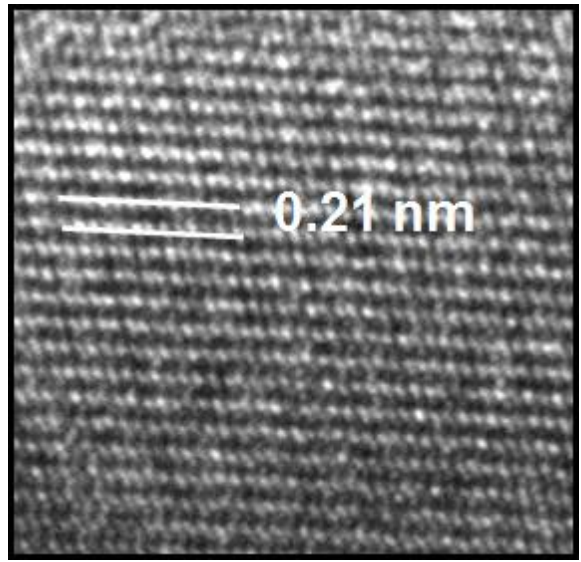

**Figure S3.** HRTEM image of bulk $Bi_2Te_3$ shows well resolved lattice fringes. The inter planar spacing is 0.21 nm.

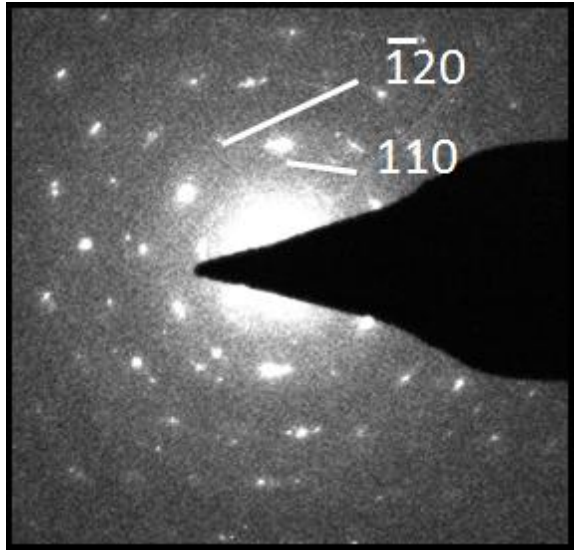

**Figure S4.** SAED pattern of bulk $Bi_2Te_3$ with well indexed diffraction spots.

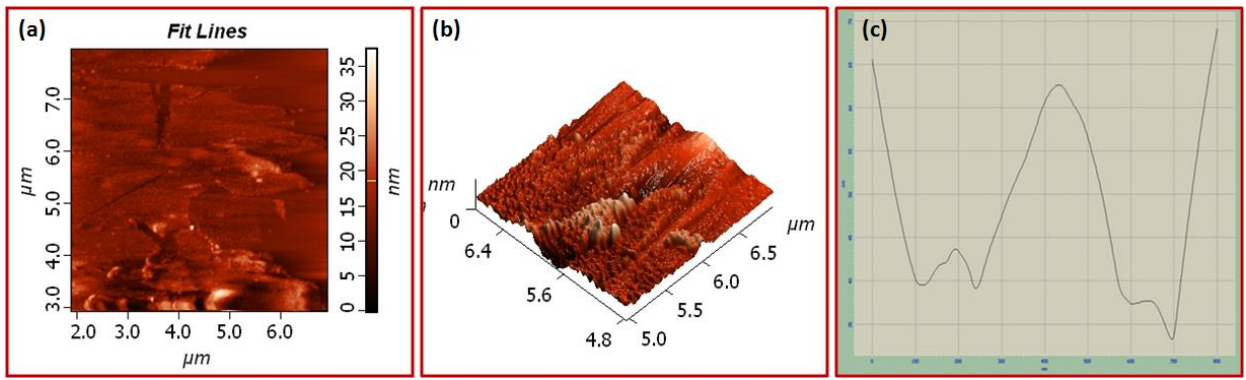

**Figure S5.** AFM micrograph of the mechanically exfoliated few layers of $Bi_2Te_3$: (a) represents the phase micrograph of few layers of $Bi_2Te_3$, where the line-scan profile is a 2d view, (b) exhibits 3d view of few layers of $Bi_2Te_3$ and (c) represents the line-scan profile, which clearly shows a thickness around (18±2) nm with wrinkle surface.

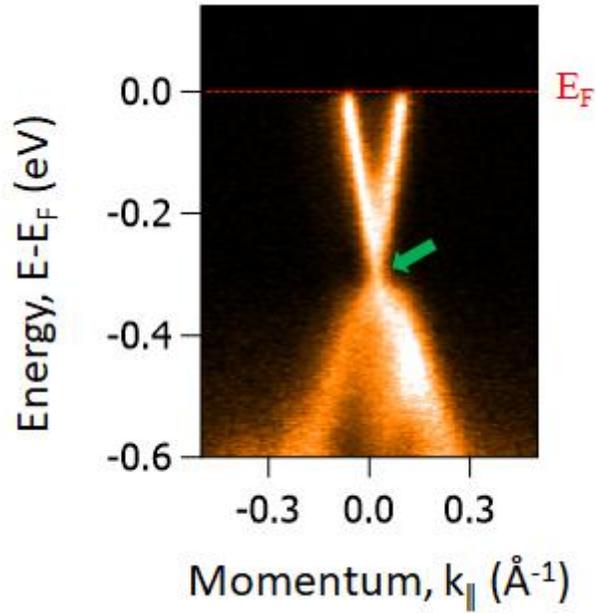

**Figure S6.** Energy distribution map (EDM) measured using ARPES depicting a Dirac point located at a binding energy of 300 meV below the Fermi level as shown by the green arrow on the EDM.

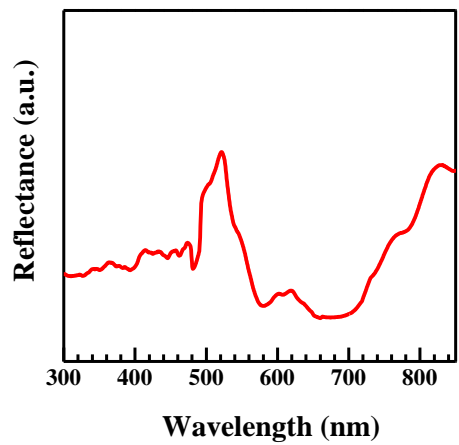

**Figure S7.** Reflectance spectrum of mechanically exfoliated few layers of $Bi_2Te_3$.

21